\documentclass[oribibl]{llncs}

\usepackage{amssymb}

\usepackage[figuresright]{rotating}
\usepackage{subfigure}

\begin{document}





\title{Feature Selection based on Machine Learning in MRIs for Hippocampal Segmentation}


\author{Sabina Tangaro\inst{1}, Nicola Amoroso\inst{1,2}, Massimo Brescia\inst{3}, Stefano Cavuoti\inst{3}, Andrea Chincarini\inst{4}, Rosangela Errico\inst{2,5}, Paolo Inglese\inst{1}, Giuseppe Longo \inst{6}, Rosalia Maglietta\inst{7}, Andrea Tateo\inst{2}, Giuseppe Riccio\inst{3}, \and Roberto Bellotti\inst{1,2}}
\institute{Istituto Nazionale di Fisica Nucleare, Sezione di Bari, Italy
\and Dipartimento Interateneo di Fisica, Universit\`a degli Studi di Bari, Italy
\and INAF - Astronomical Observatory of Capodimonte, Napoli, Italy
\and Istituto Nazionale di Fisica Nucleare, Sezione di Genova, Italy
\and Dipartimento Interateneo di Fisica, Universit\`a degli Studi di Genova, Italy
\and Dipartimento di Fisica, Universit\`a degli Studi di Napoli Federico II, Italy
\and Istituto di Studi sui Sistemi Intelligenti per l'Automazione, CNR, Bari, Italy}

\maketitle

\begin{abstract}

Neurodegenerative diseases are frequently associated with structural changes in the brain. Magnetic Resonance Imaging (MRI) scans can show
these variations and therefore  be used as a supportive feature for a number of neurodegenerative diseases. The
hippocampus has been known to be a biomarker for Alzheimer disease and other neurological and psychiatric diseases.
However, it requires accurate, robust and reproducible delineation of hippocampal structures.
Fully automatic methods are usually the voxel based approach, for each voxel a number of local features were calculated.
In this paper we compared four different techniques for feature selection from a set of $315$  features extracted for each voxel: (i) filter method based on the Kolmogorov-Smirnov test; two wrapper methods, respectively, (ii) Sequential Forward Selection and (iii) Sequential Backward Elimination; and (iv) embedded method
based on the Random Forest Classifier on a set of $10$ T1-weighted brain MRIs and tested on an independent set of $25$ subjects. The
resulting segmentations were compared with manual reference labelling. By using only $23$ feature for each voxel (sequential backward elimination) we obtained comparable state of-the-art performances with respect to the standard tool FreeSurfer \cite{freesurfer2012}.

\end{abstract}
%

%
%
%



\section{Introduction}

The analysis of medical images such as Magnetic Resonance Images (MRIs) is useful to investigate and identify the structural alterations in the brain, frequently associated with dementia or neurodegenerative diseases. In this context, the hippocampal segmentation is used to study and detect the correlation between the morphological anomalies of the hippocampus and the occurrence of the Alzheimer's disease. Hence its importance is strictly related to the early prediction of the dementia \cite{apostolova2006}.
Since the manual tracing is time-consuming and highly operator dependent, it is important to make this process as much automatic as possible.

As discussed in \cite{veronese2013}, automatic image analysis and classification methods exist, able to recognize brain anomalies at level of the single patient, which is more useful than at level of groups or categories of individuals. Nonetheless they potentially require a large amount of parameters (vector of features) to properly manage all differences and specific features of the human brain among individuals, causing the parameter space to explode in terms of complexity, redundancy and noise. To find a limited amount of features, able to recognize patterns with a sufficient level of accuracy, and without requiring a huge computational effort, would be indeed very helpful. This is especially true when the feature selection and classification are performed by machine learning techniques, since the intrinsic self-organizing selection of important features and their cross-correlation remove any potential biased interpretability of the feature space.

Several approaches have been proposed to reach different levels of automation \cite{chupin2009}. Among known methods, we quote just Morra et al. \cite{morra2010comparison,morra2008}, which suggest different automatic methods based on Support Vector Machines (SVM) and hierarchical Adaboost, by considering about $18,000$ voxel features and FreeSurfer \cite{freesurfer2012}, a standard medical software tool for the analysis of cortical and subcortical anatomy, which performs a segmentation on cortical surface streams by constructing models of boundaries among white and gray matter.

Similarly, for an automatic hippocampal segmentation we use a voxel based approach by using $315$ local features for each voxel included in a para-hippocampal region larger than the hippocampus volume. Extracting $315$ features for such a large number of voxels needs massive processing time and massive computational resources. For this reason, we consider crucial the issue of Feature Selection (FS) or reduction. The utility of feature selection is: (a) to avoid overfitting, by minimizing the dimension of the parameter space and improve model performance, i.e. prediction performance in the case of supervised classification and better cluster detection in the case of clustering, (b) to provide faster and more cost-effective models, (c) to gain a deeper insight into the underlying processes that generated the data and (d) to optimize the processing time and massive computational resource.

There is a price to be paid for this advantage. To search for a subset of relevant features introduces in fact an additional layer of complexity in the modeling task: it needs to find the optimal model parameters for the optimal feature subset, as there is no guarantee that the optimal parameters for the full input feature set are equally optimal also for the best feature subset \cite{tangaro2008,ciatto2009}.

By providing a small quantity of features, it may reduce the computational time as being proportional to the number of features. Furthermore, in some cases it allows to gain a better classification accuracy \cite{jain198239}.
Also, the reduction of the feature's number is necessary when, to train the classifier, it is available only a limited number of examples. In this regard, it is shown that, for
the same error rate, a classifier requires a training whose duration grows exponentially with the number of variables \cite{devroye1996probabilistic,masala2005,masala2007}.

Feature reduction, therefore, includes any algorithm that finds a subset of input feature set. A feature reduction capability is present also in more general methods based on transformations or combinations of the input feature set (feature extraction algorithms). An example being the well known Principal Component Analysis (PCA), which eliminates the redundancy of information by generating new features set by a combination of input features \cite{jain1997feature}.

However, the best feature selection, by preserving the original semantics of features, permits also to maintain a coherent interpretability. The main goal of this study is to exemplify and demonstrate the benefits of applying FS algorithms in hippocampus segmentation field.

\section{Materials}
\label{sec:materials}

The database used to perform the described experiments is composed by thirty-five T1-weighted whole brain MR images, and the corresponding
manually segmented bilateral hippocampi (masks). All images were acquired on a $1.0$ $T$ scanner according to MP-RAGE sequence for magnetic resonance imaging of the
brain \cite{mugler1990three, indirizzo_1, indirizzo_2}.

The images are derived from the Open Access Series of Imaging Studies (OASIS). In particular we used $35$ MP-RAGE MRI brain scans with a resolution of $1$ $mm^3$ provided in occasion of the MICCAI SATA challenge workshop 2013 \cite{indirizzomiccai}. By using this homogeneous data sample it was possible to reduce the training image sub-sample without loosing in generality and learning capabilities, giving the possibility to keep a sufficiently wide test set to perform a well-posed statistical analysis on the feature selection performances.

The image processing and classification were carried out blindly with respect to the subject status.

The first stage of our analysis chain requires an image pre-processing to standardize them both spatially and in gray intensity.
This operation is obtained by registering the images on the Montreal Neurological Institute (\textit{MNI}) standard template (ICBM152)
using 12-parameter affine-registration, and subsequent re-sampling on an isotropic grid with $1$ $mm^3$ voxel-size.

In order to reduce the computational time of the analysis, from the MRI spatially standardized, two volumes containing the left and
right hippocampus including the relevant para-hippocampal regions are extracted using a new method FAPoD (Fully automatic Algorithm
Based on Point Distribution Model) described in \cite{amoroso2012automated, Tangaro2014}.

We can then proceed with the feature extraction only in this identified region of interest : we approach a binary classification
voxel-based problem, where the categories are \textit{hippocampus} or \textit{not-hippocampus}, that is based on supervised pattern recognition systems.
The features should contain information relevant to the classification task. Since manual segmentation of the hippocampus is
based on local texture information, we adopted the related features. In the analysis presented here for each
voxel a vector whose elements represent information about position, intensity, neighboring texture \cite{lienhart2002extended}, and
local filters, was obtained.

Texture information was expressed using both Haar-like and Haralick features \cite{haralick1979statistical, morra2008validation}.

The Haralick features were calculated from the normalized gray-level co-occurrence matrices (GLCM) created on the $m \times m$ voxels projection sub-images of the volume of interest; $m$ defines the size of overlapping sliding-windows. For each voxel, values of $m$ varying
from $3$ to $9$ were used. Each element $(k,p)$ of a co-occurrence matrix indicates the probability that two voxels, separated by a specified
spatial angle and distance, have gray-level $k$ and $p$ respectively.

A subset of Haralick features is sufficient to obtain a satisfactory discrimination. To establish which of the original $14$ GLCM
Haralick features gives the best recognition rate, several preliminary recognition experiments were carried out \cite{Bellotti2006}. The resulting best
configuration has been individuated in $4$ features: energy, contrast, correlation, inverse difference moment \cite{Tangaro2014}.

Finally, the gradients calculated in different directions and at different distances were included as additional features. The best
analysis configuration, expressed by the highest values of statistical indicators (see Sec. \ref{sec:methods}), was obtained with $315$ features, described in Table \ref{features}.

By summarizing, the Knowledge Base (KB) was consisting of $35$ regions of interest (ROI) extracted from as many images, each one composed by $7910$ voxels, where each voxel is represented through a vector of $315$ features. Therefore the training set, including $10$ randomly selected images, was formed by a total of $79100 \times 315$ entries. In quantitative terms it can be considered a sufficiently wide data set, qualitatively able to cover all feature types, needed to perform a complete training, avoiding the useless redundancy of information not needed by machine learning methods \cite{Brescia2014}, and leaving a sufficiently large amount of samples to be dedicated to the test sessions.

\begin{table}
\begin{center}
\begin{tabular}{|l|l|}
  \hline
  number & description \\ \hline
  $1$ & position \\
  $1$ & grey level \\
  $66$ & Haralick features for mask $3\times3$  \\
  $66$ & Haralick features for mask $5\times5$  \\
  $66$ & Haralick features for mask $7\times7$ \\
  $66$ & Haralick features for mask $9\times9$ \\
  $49$ & Haar-like 3D features \\
  \hline
\end{tabular}
\end{center}
\caption{The $315$ features extracted from the 3D MRI images. Of each group of $66$ Haralick features, $13$ are the gradients along the $13$ diagonals, $5$ the principal moments, and the rest are the three sets of $16$ textural features, one set for each plane of the voxels. The gradients for each voxel are measured in all directions at one voxel distance and the relative 3D positions are included as features.}\label{features}
\end{table}

\section{Methods}
\label{sec:methods}

The FS techniques are usually counted in three categories, based on their internal combination between the selection and classification of the reduced parameter space. These categories are respectively named as wrapper, filter and embedded methods \cite{saeys2007review}.

\textit{Filter} method is a technique based on the measurement of the importance of each single feature of the given parameter space \cite{sanchez2007}. The selected features are the most relevant to obtain a correct classification. This technique include methods suitable for high-dimensional datasets, since they are computationally fast. Furthermore, they are independent from the classification algorithm and therefore their results can be used for all types of classifier. However, since each feature is considered separately from the others, their positive contribution based on the combined effect is neglected. The filter method used in our analysis is based on the Kolmogorov-Smirnov (K-S) test.

\textit{Wrapper} methods basically integrate the two aspects of the workflow, i.e the model hypothesis and feature search \cite{kohavi1997}. This procedure involves the generation and evaluation of various subsets of features. Every generated feature subset is associated to a classification criterion (hence the name \textit{wrapper}). Since the number of all possible feature subsets grows exponentially with the size of the dataset, some search heuristics can be adopted to reduce drastically the number of operations. They can be grouped into \textit{deterministic} and \textit{randomized} search methods. The advantage of these methods is the intrinsic best interaction among selected features and their classifiers, but with the downside to have a high computational cost and the risk of overfitting. The wrapper methods used in our analysis are respectively, sequential forward selection (SFS)  and sequential backward elimination (SBE).

Finally, in \textit{embedded} methods the optimal feature subset search is directly nested into the classifier algorithm \cite{vapnik1998}. Such techniques can be interpreted in terms of a search within a combined parameter space, by mixing features and hypotheses. Analogously to wrapper methods, they include the interaction with classification algorithm, but in a faster way. The embedded method used in our analysis is based on the Random Forest Classifier.

To recap, in our FS analysis we used:
\begin{itemize}
  \item Univariate filter method: Kolmogorov-Smirnov,
  \item Deterministic wrapper methods: sequential forward selection
  (SFS) and sequential backward elimination (SBE);
  \item Embedded method: Random Forest.
\end{itemize}

In addition, we have also used the PCA \cite{abdi2010}, being one of the most widely adopted feature reduction techniques, for comparison.

To estimate the goodness of the selected feature group we used the N\"{a}ive Bayes Classifier \cite{rish2001empirical}, based on the simplified hypothesis that all attributes describing a specific instance on data are conditionally independent among themselves.

The FS analysis was performed in the $5$-fold cross validation on $10$ of $35$ images in the database. The goodness of the selected group was tested on the remaining $25$ images.
As already discussed in Sec.~\ref{sec:materials}, the selected training and test rates were considered sufficiently wide to ensure a well-posed training and the post-processing statistical evaluation.

The k-fold cross validation is a technique able to avoid overfitting on data and is able to improve the generalization performance of the machine learning model. In this way, validation can be implicitly performed during training, by enabling at setup the standard leave-one-out k-fold cross validation mechanism \cite{geisser1975}. The automatized process of the cross-validation consists in performing $k$ different training runs with the following procedure: (i) splitting of the training set into $k$ random subsets, each one composed by the same percentage of the data set (depending on the $k$ choice); (ii) at each run the remaining part of the data set is used for training and the excluded percentage for validation. While avoiding overfitting, the k-fold cross validation leads to an increase of the execution time estimable around $k-1$ times the total number of runs.

Furthermore, the combination of the Bayes rule with the above simplified assumption has a positive impact on the model complexity and its computational time. In particular, the latter property pushed us to choose this model as embedded classifier for the feature selection problem.

The agreement between an automated segmentation estimate and a manual segmentation can be assessed using overlap measures. A number of measures are available: (a) Dice Index \cite{Tangaro2014}, \cite{dice1945measures}; (b) efficiency; (c) purity of a class; (d) completeness of a class; (e) contamination of a class.

At the base of the statistical indicators adopted, there is the commonly known \textit{confusion matrix}, which can be used to easily visualize the
classification performance \cite{provost1998}: each column of the matrix represents the instances in a predicted class, while each row represents the instances
in the real class. One benefit of a confusion matrix is the simple way in which it allows to see whether the system is mixing different
classes or not.

We remark here that we were mostly interested to the feature analysis related to the classification of the \textit{hippocampus} class voxels. Therefore we considered as particularly relevant the Dice index, usually referred to the \textit{true positive} class ($N_{AA}$ in our confusion matrix), which in our case correspond properly to \textit{hippocampus} class. Since, by definition, the Dice index does not take the true negative rate into account, the rate of \textit{not-hippocampus} voxels is not involved within this indicator. A statistical evaluation of this latter class, corresponding to the background voxels, has been primarily included for completeness and for coherency with the full confusion matrix representation. The highest relevance given to the \textit{hippocampus} class analysis represents also a common evaluation criterion in such context \cite{morra2008}.

In terms of binary classification, we were more interested to perform a feature selection analysis, rather than to improve the classification performances. Therefore we imposed a standard classification threshold to $0.5$ at the beginning of the experiments and maintained unchanged all over the entire described process, by considering it as sufficient for our specific purposes.

More specifically, for a generic two-class confusion matrix,

\begin{eqnarray}
  \begin{array}{c|ccc}
            &            & $OUTPUT$\\ \hline
            &    -       &$Class A$      & $Class B$ \\
    $TARGET$  &$Class A$    & N_{AA}              & N_{AB} \\
            &$Class B$    & N_{BA}              & N_{BB} \\
  \end{array} \nonumber
  \end{eqnarray}

\noindent we then use its entries to define the following statistical quantities:

\begin{itemize}
\item \underline{total efficiency}: $te$. Defined as the ratio between the number of correctly classified objects and the total number of objects in the data set.
In our confusion matrix example it would be:
    \begin{eqnarray}
    te=\frac{N_{AA} + N_{BB}}{N_{AA} + N_{AB} + N_{BA} + N_{BB}} \nonumber
    \end{eqnarray}
\item \underline{purity of a class}: $pcN$. Defined as the ratio between the number of correctly classified objects of a class and the number of objects classified
in that class. In our confusion matrix example it would be:
    \begin{eqnarray}
    pcA=\frac{N_{AA}}{N_{AA}+N_{BA}} \nonumber
    \end{eqnarray}
    \begin{eqnarray}
    pcB=\frac{N_{BB}}{N_{AB}+N_{BB}} \nonumber
    \end{eqnarray}
\item \underline{completeness of a class}: $cmpN$. Defined as the ratio between the number of correctly classified objects in that class and the total number of
objects of that class in the data set. In our confusion matrix example it would be:
    \begin{eqnarray}
    cmpA=\frac{N_{AA}}{N_{AA}+N_{AB}} \nonumber
    \end{eqnarray}
    \begin{eqnarray}
    cmpB=\frac{N_{BB}}{N_{BA}+N_{BB}} \nonumber
    \end{eqnarray}
\item \underline{contamination of a class}: $cntN$. It is the dual of the purity, namely it is the ratio between the misclassified objects in a class and the number
of objects classified in that class, in our confusion matrix example will be:
    \begin{eqnarray}
    cntA=1-pcA=\frac{N_{BA}}{N_{AA}+N_{BA}} \nonumber
    \end{eqnarray}
    \begin{eqnarray}
    cntB=1-pcB=\frac{N_{AB}}{N_{AB}+N_{BB}} \nonumber
    \end{eqnarray}
\item \underline{Dice index}: $Dice$. Known also with the name of $F_1 score$, it is a frequent measure used in binary classification, which could be considered as a weighted average of the purity and completeness, reaching its best value at $1$ and the worst at $0$. By referring to our notation, we have the Dice defined as:
    \begin{eqnarray}
    Dice=2\cdot\frac{pcA*cmpA}{pcA+cmpA}= 2\cdot\frac{N_{AA}}{2N_{AA}+N_{BA}+N_{AB}}\nonumber
    \end{eqnarray}
\end{itemize}

\section{Results}


By using N\"{a}ive Bayes Classifier on all $315$ input features, the goodness is estimated in $5$-fold cross validation on $10$ images.
The results in terms of the statistics derived from the confusion matrix, are shown in Table \ref{man_88_nap} and the Dice index is $0.60 \pm 0.04$.

\begin{table}[H]
\begin{center}
         \begin{tabular}{|c|c|c|c|}
         \hline
         \textbf{$\textbf{315}$} & \textbf{Completeness} & \textbf{Purity } & \textbf{Contamination} \\
         \textbf{input features} & \textbf{of a class} & \textbf{of a class} & \textbf{of a class} \\
         \hline
         \textbf{Hippocampus}      & $79\%$  & $62\%$ & $38\%$ \\
         \hline
         \textbf{Not Hippocampus}      & $63\%$  & $80\%$ & $20\%$ \\
         \hline
         \hline
         \textbf{Efficiency} & \multicolumn{3}{|c|}{$70\%$} \\
         \hline
         \end{tabular}
  \end{center}
          \caption{Classification result on all $315$ input features using N\"{a}ive Bayes Classifier in $5$-fold cross validation based on confusion matrix.}\label{man_88_nap}
\end{table}


The PCA applied to $315$ input features returns the principal components (PCs) ordered by the amount of information they convey. The percentage of information contained in the first $98$ PCs and in the first $197$ PCs are respectively $90\%$ and $99\%$.

Since our goal was to reduce the feature retaining the goodness in the classification, we considered the first $197$ PCs containing $99.0\%$ of the information. The results obtained are shown in Table \ref{man_4_nap} and the Dice index is $0.62 \pm 0.07$. As above, we used the N\"{a}ive Bayes Classifier in $5$-fold cross validation.

\begin{table}[H]
\begin{center}
         \begin{tabular}{|c|c|c|c|}
         \hline
         \textbf{$197$ } & \textbf{Completeness} & \textbf{Purity } & \textbf{Contamination} \\
         \textbf{PCs} & \textbf{of a class} & \textbf{of a class} & \textbf{of a class} \\
         \hline
         \textbf{Hippocampus}      & $60\%$  & $68\%$ & $32\%$ \\
         \hline
         \textbf{Not Hippocampus}      & $78\%$  & $72\%$ & $28\%$ \\
         \hline
         \hline
         \textbf{Efficiency} & \multicolumn{3}{|c|}{$71\%$} \\
         \hline
         \end{tabular}
  \end{center}
          \caption{Classification result on the first $197$ PCs using N\"{a}ive Bayes Classifier using in $5$-fold
          cross validation based on confusion matrix.}\label{man_4_nap}
\end{table}

Compared to the use of all $315$ original features, the values obtained with $197$ PCs are on average $6\%$ points lower in terms of
Dice Index. Therefore, to avoid loss of information, we considered all $315$ PCs. The results are reported in table \ref{man_5_nap} and the Dice index is $0.63 \pm 0.03$.

\begin{table}[H]
\begin{center}
         \begin{tabular}{|c|c|c|c|}
         \hline
         \textbf{$315$ } & \textbf{Completeness} & \textbf{Purity } & \textbf{Contamination} \\
         \textbf{PCs} & \textbf{of a class} & \textbf{of a class} & \textbf{of a class} \\
         \hline
         \textbf{Hippocampus}      & $86\%$  & $51\%$ & $49\%$ \\
         \hline
         \textbf{Not Hippocampus}      & $36\%$  & $78\%$ & $22\%$ \\
         \hline
         \hline
         \textbf{Efficiency} & \multicolumn{3}{|c|}{$58\%$} \\
         \hline
         \end{tabular}
  \end{center}
          \caption{Classification result on all $315$ PCs using N\"{a}ive Bayes Classifier using in
          $5$-fold cross validation based on confusion matrix.}\label{man_5_nap}
\end{table}

Even using all the PCs, the result was $5\%$ points lower in terms of Dice Index. This result confirms what already found by Golland et al. in \cite{golland2005detection}, i.e. that
the selection of large-variance features performed by the PCA is not specifically suited for segmentation problems.

\subsection{Kolmogorov-Smirnov analysis}

The (K-S) test provides an estimate of how much two distributions are related to each other. The K-S test allowed us to select only the features which have a correlation between the two \textit{hippocampus} and \textit{not-hippocampus} classes less than $5\%$, resulting in a total of $57$ features.

As above, we used the N\"{a}ive Bayes Classifier in $5$-fold cross validation. The results obtained are shown in table \ref{man_3_nap} and the Dice index is $0.67 \pm 0.04$.

\begin{table}[H]
\begin{center}
         \begin{tabular}{|c|c|c|c|}
         \hline
         \textbf{$57$ features} & \textbf{Completeness} & \textbf{Purity } & \textbf{Contamination} \\
         \textbf{Kolmogorov-Smirnov} & \textbf{of a class} & \textbf{of a class} & \textbf{of a class} \\
         \hline
         \textbf{Hippocampus}      & $84\%$  & $57\%$ & $43\%$ \\
         \hline
         \textbf{Not Hippocampus}      & $52\%$  & $81\%$ & $19\%$ \\
         \hline
         \hline
         \textbf{Efficiency} & \multicolumn{3}{|c|}{$66\%$} \\
         \hline
         \end{tabular}
  \end{center}
          \caption{Classification result on $57$ features selected trough Kolmogorov-Smirnov test
          using N\"{a}ive Bayes Classifier using in $5$-fold cross validation based on confusion matrix.}\label{man_3_nap}
\end{table}

The K-S test results are comparable with the original parameters space based on $315$ features.

\subsection{Sequential forward selection and backward elimination}

\begin{figure}[H]
\centering
\subfigure[]
{\includegraphics[scale=0.3]{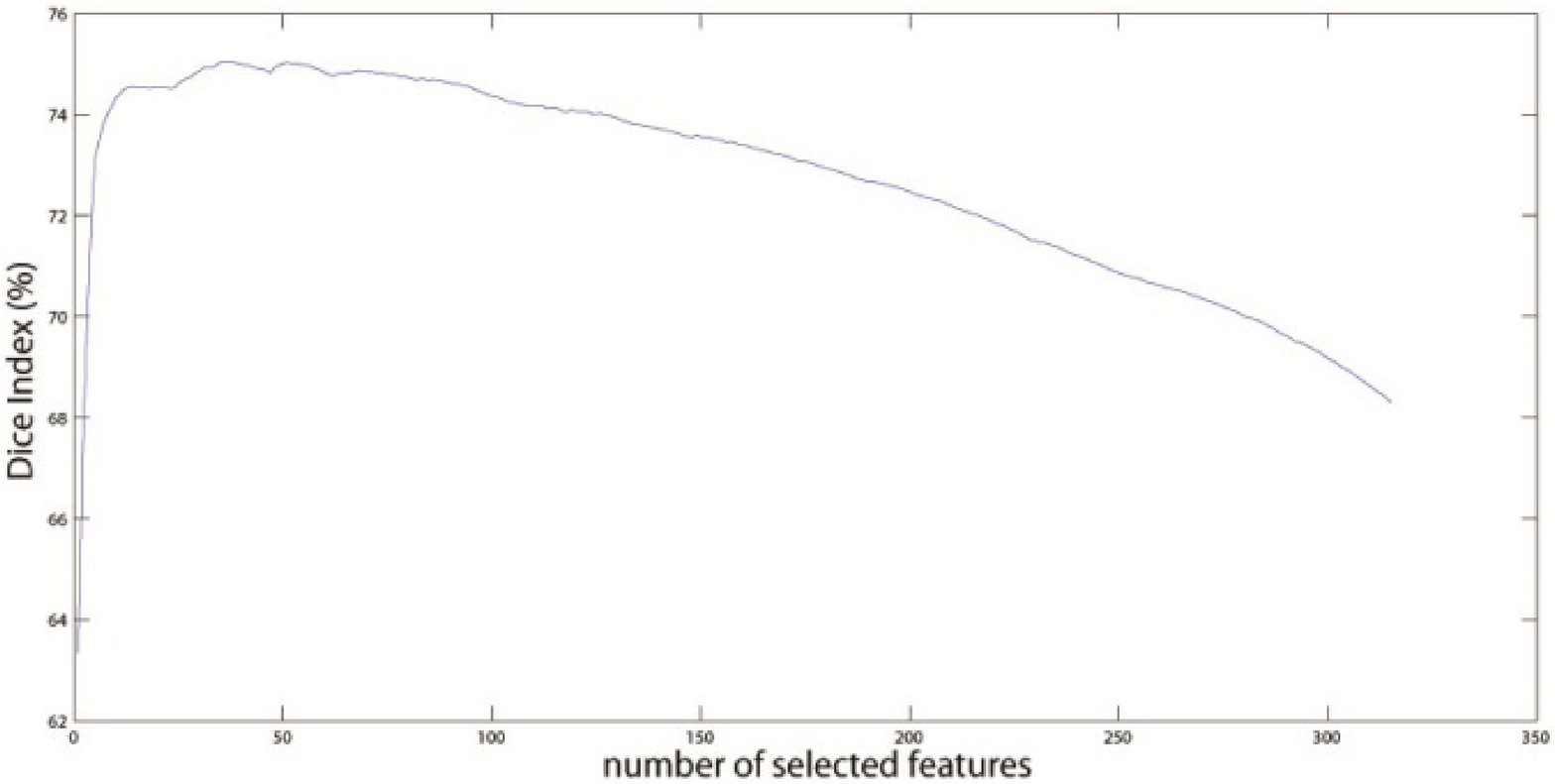}}
\subfigure[]
 {\includegraphics[scale=0.3]{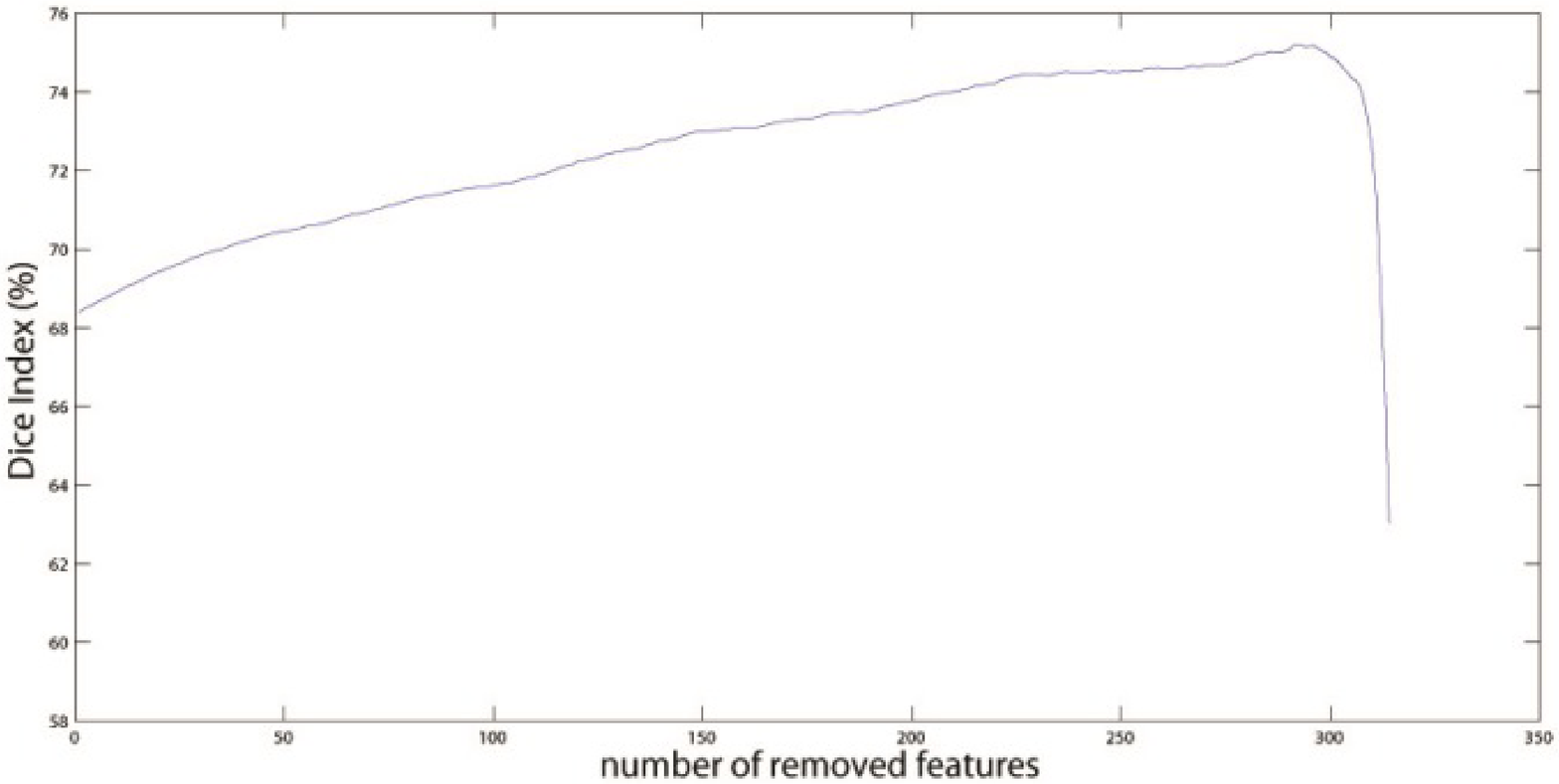}}
 \caption{Best Dice Index of all the possible combinations of the relevant step in (a) sequential forward selection and in (b) sequential backward elimination methods.}\label{ForwardSelect}
\end{figure}

The two FS methods belonging to the wrapper category experimented in our case were SFS and SBE.
In Figure \ref{ForwardSelect}(a) on the ordinate axis it is shown the top value of Dice Index achieved between all possible combinations
related to the reference step depicted on the horizontal axis. At each step, the feature achieving the best performance is chosen,
when used in combination with the selected features in the previous step. The step number coincides with the number of selected
features (SFS).

In Figure \ref{ForwardSelect}(b) on the ordinate axis it is shown the top value of Dice Index achieved between all possible combinations related to the reference step depicted on the horizontal axis. At each step it is removed the feature without which the best performances are obtained. The step number coincides with the number of eliminated features (SBE).

We observe that the SFS method reaches its highest Dice Index, $0.75$, at the step $36$. So it means that the best performance, using the N\"{a}ive Bayes Classifier, is obtained with
only $36$ selected features, listed in Tab.~\ref{sfs_features}.

The SBE method obtains, its highest Dice Index, $0.75$ at the step $292$. Therefore the best performance, evaluated with the N\"{a}ive Bayes Classifier, is obtained by using the remaining $23$ features (i.e. $315-292$), listed in Tab.~\ref{sbe_features}.

Tables \ref{man_7_nap_1} (related Dice index is $0.75 \pm 0.03$) and \ref{man_8_nap_1} (related Dice index is $0.75 \pm 0.02$), respectively, show the relative performance of the
peak value in Figure \ref{ForwardSelect}.

\begin{table}[H]
\begin{center}
         \begin{tabular}{|c|c|c|c|}
         \hline
         \textbf{$36$ features} & \textbf{Completeness} & \textbf{Purity } & \textbf{Contamination} \\
         \textbf{Forward Selection} & \textbf{of a class} & \textbf{of a class} & \textbf{of a class} \\
         \hline
         \textbf{Hippocampus}      & $82\%$  & $70\%$ & $30\%$ \\
         \hline
         \textbf{Not Hippocampus}      & $73\%$  & $84\%$ & $16\%$ \\
         \hline
         \hline
         \textbf{Efficiency} & \multicolumn{3}{|c|}{$77\%$} \\
         \hline
         \end{tabular}
  \end{center}
          \caption{Classification result on $36$ features selected trough forward selection method using N\"{a}ive Bayes Classifier in $5$-fold cross validation based on confusion matrix.}\label{man_7_nap_1}
\end{table}

\begin{figure}[H]
\begin{center}
     \includegraphics[scale=0.50]{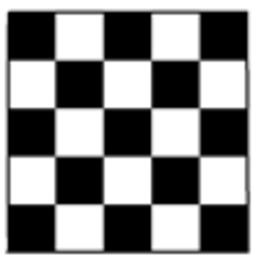}
     \includegraphics[scale=0.50]{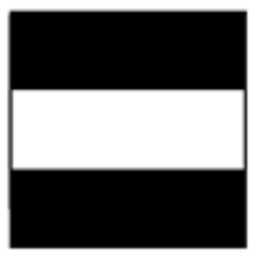}
\end{center}
     \caption{Haar-like template types $1$ (left) and $2$ (right) used in the experiments.}\label{templates}
 \end{figure}

\begin{figure}[H]
\begin{center}
     \includegraphics[scale=1]{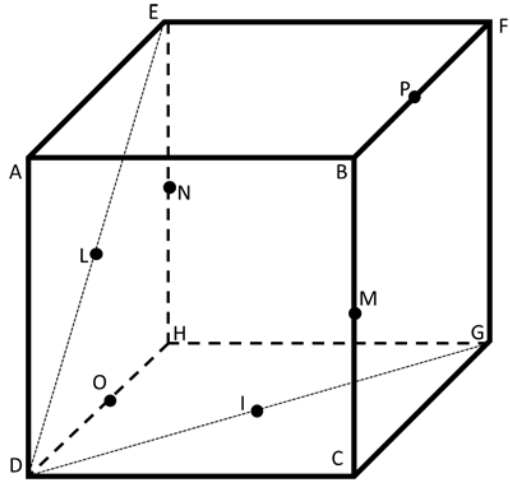}
\end{center}
     \caption{Representation of a generic cubic mask used for calculating the gradient features. The labeled points are either the vertexes of the cube or the median points of the segments.}\label{cube}
 \end{figure}

\begin{table}[H]
\begin{center}
         \begin{tabular}{|c|c|c|c|c|c|c|}
         \hline
         \textbf{$36$ features}          & \multicolumn{3}{|c|}{\textbf{Haralick}} &\textbf{Haar-like} & \multicolumn{2}{|c|}{\textbf{Statistical}}  \\
         \textbf{Forward Selection}      &\textbf{Orientation} &\textbf{Coordinate} &\textbf{Msize}  &\textbf{Type}  &\textbf{Msize} &\textbf{Entry} \\
         \hline
         \textbf{contrast*}              &135         &Y                 &3                  &                   &                     & \\
         \hline
         \textbf{gradient*}              &            &                  &                   &                   &5                    &$\overline{EC}$ \\
         \hline
         \textbf{correlation}            &135         &X                 &3                  &                   &                     & \\
         \hline
         \textbf{position*}              &            &                  &                   &                   &                     &coords \\
         \hline
         \textbf{norm. gray level*}      &            &                  &                   &                   &                     &value \\
         \hline
         \textbf{correlation*}           &45          &X                 &5                  &                   &                     & \\
         \hline
         \textbf{gradient*}              &            &                  &                   &                   &5                    &$\overline{DF}$ \\
         \hline
         \textbf{correlation*}           &90          &Y                 &9                  &                   &                     & \\
         \hline
         \textbf{correlation}            &45          &Y                 &7                  &                   &                     & \\
         \hline
         \textbf{skewness*}              &            &                  &                   &                   &7                    & \\
         \hline
         \textbf{homogeneity*}           &90          &X                 &9                  &                   &                     & \\
         \hline
         \textbf{correlation}            &0           &Y                 &5                  &                   &                     & \\
         \hline
         \textbf{correlation}            &90          &Z                 &5                  &                   &                     & \\
         \hline
         \textbf{correlation*}           &45          &X                 &3                  &                   &                     & \\
         \hline
         \textbf{correlation}            &135         &Z                 &9                  &                   &                     & \\
         \hline
         \textbf{correlation}            &90          &Y                 &5                  &                   &                     & \\
         \hline
         \textbf{correlation}            &135         &Z                 &5                  &                   &                     & \\
         \hline
         \textbf{correlation}            &0           &Z                 &7                  &                   &                     & \\
         \hline
         \textbf{correlation}            &90          &Z                 &7                  &                   &                     & \\
         \hline
         \textbf{correlation}            &90          &Z                 &9                  &                   &                     & \\
         \hline
         \textbf{correlation}            &0           &Y                 &3                  &                   &                     & \\
         \hline
         \textbf{correlation}            &135         &X                 &3                  &                   &                     & \\
         \hline
         \textbf{correlation}            &0           &Z                 &9                  &                   &                     & \\
         \hline
         \textbf{template*}              &            &                  &                   &1                  &                     & \\
         \hline
         \textbf{skewness*}              &            &                  &                   &                   &5                    & \\
         \hline
         \textbf{correlation}            &90          &Z                 &3                  &                   &                     & \\
         \hline
         \textbf{correlation}            &45          &X                 &5                  &                   &                     & \\
         \hline
         \textbf{gradient}               &            &                  &                   &                   &3                    &$\overline{MN}$ \\
         \hline
         \textbf{template}               &            &                  &                   &2                  &                     & \\
         \hline
         \textbf{correlation*}           &45          &X                 &9                  &                   &                     & \\
         \hline
         \textbf{correlation}            &45          &Y                 &5                  &                   &                     & \\
         \hline
         \textbf{correlation}            &90          &Y                 &7                  &                   &                     & \\
         \hline
         \textbf{correlation}            &45          &Z                 &5                  &                   &                     & \\
         \hline
         \textbf{gradient}               &            &                  &                   &                   &9                    &$\overline{DF}$ \\
         \hline
         \textbf{homogeneity}            &0           &Z                 &9                  &                   &                     & \\
         \hline
         \textbf{correlation}            &0           &Y                 &9                  &                   &                     & \\
         \hline
         \end{tabular}
  \end{center}
          \caption{Details of the $36$ features resulting by the forward selection method using N\"{a}ive Bayes Classifier. The asterisk indicates the entries also present in the list of $23$ SBE features. For Haralick features, the orientation in degrees, reference coordinate and the size of the cubic mask used are reported. In case of Haar-like features, the entry value indicates the template type used (see Fig.~\ref{templates}). For statistical/positional kind there are listed the size of the cubic mask used or the self-explained value, depending on the specific feature type. In particular for gradients, the column Entry indicates the segment of the reference diagonal as shown in Fig.~\ref{cube}. All the features are listed in top-down order of their inclusion during the SFS procedure execution.}\label{sfs_features}
\end{table}

\begin{table}[H]
\begin{center}
         \begin{tabular}{|c|c|c|c|}
         \hline
         \textbf{$23$ features} & \textbf{Completeness} & \textbf{Purity } & \textbf{Contamination} \\
         \textbf{Backward Elimination} & \textbf{of a class} & \textbf{of a class} & \textbf{of a class} \\
         \hline
         \textbf{Hippocampus}      & $83\%$  & $70\%$ & $30\%$ \\
         \hline
         \textbf{Not Hippocampus}      & $73\%$  & $85\%$ & $15\%$ \\
         \hline
         \hline
         \textbf{Efficiency} & \multicolumn{3}{|c|}{$77\%$} \\
         \hline
         \end{tabular}
  \end{center}
          \caption{Classification result on $23$ features selected trough backward elimination method using N\"{a}ive Bayes Classifier in $5$-fold cross validation based on confusion matrix.}\label{man_8_nap_1}
\end{table}

\begin{table}[H]
\begin{center}
         \begin{tabular}{|c|c|c|c|c|c|c|}
         \hline
         \textbf{$23$ features}          & \multicolumn{3}{|c|}{\textbf{Haralick}} &\textbf{Haar-like} & \multicolumn{2}{|c|}{\textbf{Statistical}}  \\
         \textbf{Backward Elimination}   &\textbf{Orientation} &\textbf{Coordinate}    &\textbf{Msize}    &\textbf{Type}    &\textbf{Msize}  &\textbf{Entry} \\
         \hline
         \textbf{position*}              &            &                  &                   &                   &                     &coords \\
         \hline
         \textbf{norm. gray level*}      &            &                  &                   &                   &                     &value \\
         \hline
         \textbf{correlation}            &0           &Y                 &7                  &                   &                     & \\
         \hline
         \textbf{correlation*}           &45          &X                 &3                  &                   &                     & \\
         \hline
         \textbf{correlation*}           &45          &X                 &5                  &                   &                     & \\
         \hline
         \textbf{correlation}            &45          &X                 &7                  &                   &                     & \\
         \hline
         \textbf{correlation*}           &45          &X                 &9                  &                   &                     & \\
         \hline
         \textbf{correlation}            &45          &Y                 &9                  &                   &                     & \\
         \hline
         \textbf{correlation}            &45          &Y                 &5                  &                   &                     & \\
         \hline
         \textbf{correlation*}           &90          &Y                 &9                  &                   &                     & \\
         \hline
         \textbf{homogeneity}            &135         &Z                 &3                  &                   &                     & \\
         \hline
         \textbf{gradient*}              &            &                  &                   &                   &5                    &$\overline{DF}$ \\
         \hline
         \textbf{gradient}               &            &                  &                   &                   &7                    &$\overline{DF}$ \\
         \hline
         \textbf{contrast*}              &135         &Y                 &3                  &                   &                     & \\
         \hline
         \textbf{gradient}               &            &                  &                   &                   &9                    &$\overline{OP}$ \\
         \hline
         \textbf{homogeneity*}           &90          &X                 &9                  &                   &                     & \\
         \hline
         \textbf{gradient}               &            &                  &                   &                   &3                    &$\overline{BH}$ \\
         \hline
         \textbf{skewness*}              &            &                  &                   &                   &7                    & \\
         \hline
         \textbf{gradient*}              &            &                  &                   &                   &5                    &$\overline{EC}$ \\
         \hline
         \textbf{gradient}               &            &                  &                   &                   &3                    &$\overline{IL}$ \\
         \hline
         \textbf{template*}              &            &                  &                   &1                  &                     & \\
         \hline
         \textbf{skewness*}              &            &                  &                   &                   &5                    & \\
         \hline
         \textbf{gradient}               &            &                  &                   &                   &5                    &$\overline{MN}$ \\
         \hline
         \end{tabular}
  \end{center}
          \caption{Details of the $23$ features resulting by the backward elimination method using N\"{a}ive Bayes Classifier. The asterisk indicates the entries also present in the list of $36$ SFS features. For Haralick features, the orientation in degrees, reference coordinate and the size of the cubic mask used are reported. In case of Haar-like features, the entry value indicates the template type used (see Fig.~\ref{templates}). For statistical/positional kind there are listed the size of the cubic mask used and/or the self-explained value, depending on the specific feature type. In particular for gradients, the column Entry indicates the segment of the reference diagonal as shown in Fig.~\ref{cube}.}\label{sbe_features}
\end{table}

\subsection{Random Forest analysis}

The Random Forest classification methodology allowed us to estimate the feature importance \cite{breiman2001random}. To select the best subset we have performed a study of classification with cross validation procedure based on the N\"{a}ive Bayes Classifier, varying the threshold on the feature importance index. The optimal threshold was related to the maximum Dice Index value and achieved with $222$ features. Also in this case we used the N\"{a}ive Bayes Classifier in $5$-fold cross validation to evaluate the features selected by the Random Forest. The result obtained is shown in table \ref{man_7_nap} and the Dice index is $0.69 \pm 0.04$.

\begin{table}[H]
\begin{center}
         \begin{tabular}{|c|c|c|c|}
         \hline
         \textbf{$222$ features} & \textbf{Completeness} & \textbf{Purity } & \textbf{Contamination} \\
         \textbf{Random Forest} & \textbf{of a class} & \textbf{of a class} & \textbf{of a class} \\
         \hline
         \textbf{Hippocampus}      & $80\%$  & $62\%$ & $38\%$ \\
         \hline
         \textbf{Not Hippocampus}      & $62\%$  & $80\%$ & $20\%$ \\
         \hline
         \hline
         \textbf{Efficiency} & \multicolumn{3}{|c|}{$70\%$} \\
         \hline
         \end{tabular}
  \end{center}
          \caption{Classification result on $222$ features selected trough Random Forest method using N\"{a}ive Bayes Classifier in $5$-fold cross validation based on confusion matrix.}\label{man_7_nap}
\end{table}

\subsection{Random selection test}

\begin{figure}[H]
\begin{center}
  \includegraphics[scale=0.45]{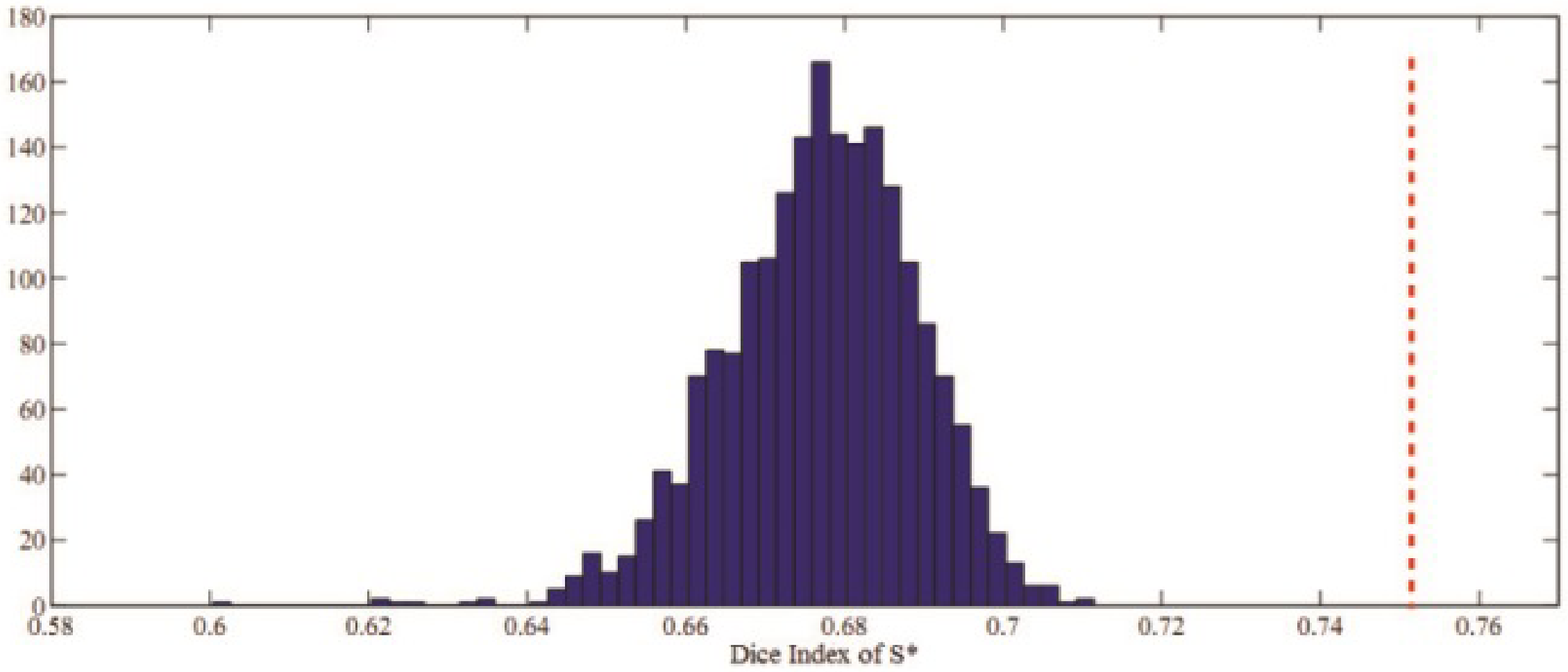}\\
   \end{center}
   \caption{Distribution of $2000$ random Dice values compared with true Dice (shown with the dashed red line) concerning to $36$ features obtained by the sequential forward selection .}\label{random_1}
\end{figure}

Furthermore we performed an additional group of tests to evaluate whether randomly selected samples of $36$ features among the original $315$, might lead to Dice
indexes greater than or comparable with the Dice value obtained with SFS ($0.75$). To do so, we estimate the empirical probability density function of
Dice under the null hypothesis that any set S* of $36$ features provides a Dice value greater than or equal to the true Dice in predicting whether a voxel belongs to hippocampus or
not.
To test this hypothesis, $2000$ sets S* were generated, each composed of $36$ features randomly drawn from the ones available and the corresponding Dice values were evaluated. The obtained results are shown in Figure \ref{random_1}.

\section{Discussion and Conclusion}

Main goal of this work was to verify the possibility to reduce the number of required voxel features without loosing or better by enhancing the classification performances. Moreover the reduction of the number of voxel features could also improve the computational efficiency of the classification.

As clearly resulted from a recent review, \cite{veronese2013}, by now the feature selection has to be considered an essential step within the field of neuroimaging approached by the machine learning paradigm. Its importance is also invariant to the specific technique used to extract and codify the features from MRIs regions of interest, whether it is based on standard n-dimensional feature vectors or on pairwise dissimilarity representation. In the present work we investigated the application of several feature selection methods.

\begin{figure}[H]
\begin{center}
     \includegraphics[scale=0.50]{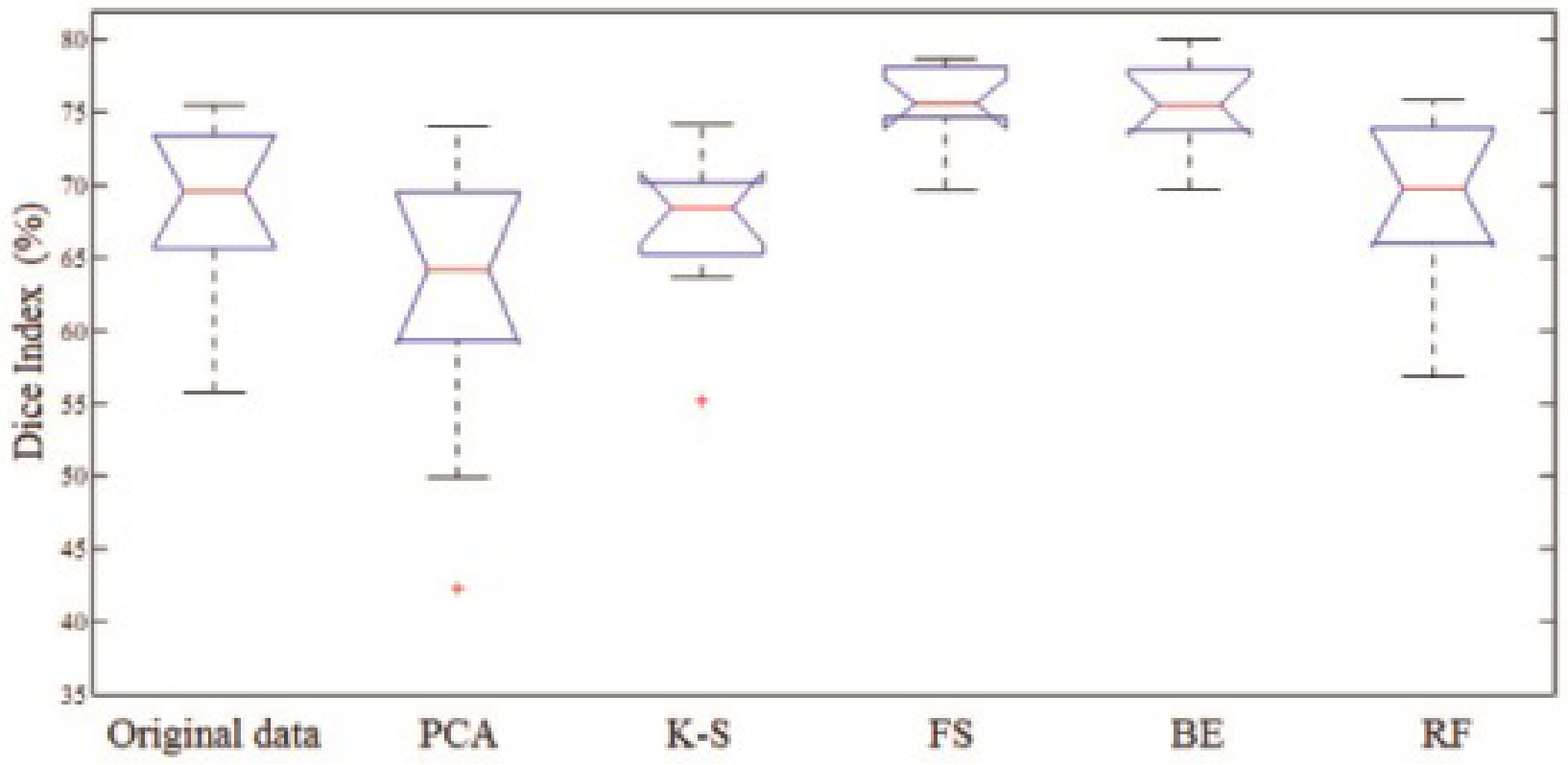}
\end{center}
     \caption{Dice Index comparison for the following methods: Original dataset ($315$ for each voxel); PCA ($197$ selected features); K-S test ($57$ selected
features); SFS ($36$ selected features); SBE ($23$ selected features); Random Forest ($222$ selected features). Boxes have lines at the lower quartile, median, and upper quartile values, with whiskers extending to 1.5 times the inter-quartile range. Outliers are indicated by a plus sign.}\label{box_1}
 \end{figure}

The results obtained using different approaches are summarized in Table \ref{tab_rias} and in Figure \ref{box_1}. We observe that using this two selected subsets it is
possible to obtain higher performances than using the entire input dataset.

\begin{table}[H]
\begin{center}
      \begin{tabular}{|c|c|c|}
         \hline

              \textbf{Method}           &     \textbf{Size selected group}     &  \textbf{Dice Index } \\
         \hline
         original dataset &  $315$  & $0.69 \pm  0.04$   \\
         \hline
         PCA selection & $197$   & $0.62 \pm 0.07$ \\
         \hline
         K-S selection & $57$    & $0.67 \pm 0.04$ \\
         \hline
         Forward Selection & $36$   & $0.75\pm 0.02$ \\
         \hline
         Backward Elimination & $23$   & $0.75 \pm0.02$ \\
         \hline
         Random Forest & $222$    & $0.69 \pm  0.04$ \\
         \hline
      \end{tabular}
\end{center}
          \caption{For each implemented method size selected group, mean Dice Index (evaluated using N\"{a}ive Bayes Classifier) and related $\sigma$ are shown.}\label{tab_rias}
\end{table}

By considering the percentage of random Dice values bigger than the best one with respect to the total number of random extractions, such value is zero. But, as it can be seen in Figure \ref{random_1}, in many cases it appears to obtain better performances by randomly extracting the feature sample, than to consider the complete set of $315$ features.

Among the FS approaches presented in this work, the SFS and SBE show better performances.

We would underline that the results shown in Figure \ref{box_1} have to be mainly interpreted as a comparison among the different methods of feature selection. What has to be stressed is that the performances are influenced by the feature information content as well as the image enhancement techniques employed. A quite simple method, such as the N\"{a}ive Bayes Classifier, is able to reach state of the art performances when preceded by a selection analysis on the feature space. A more detailed study of the classification methods and of the post-processing technique which can be used to improve performances are presented in other studies \cite{tangaro2013,maglietta2013}.

To test the goodness of the best feature selection methods presented in this paper we used the two selected sets formed, respectively, by $36$ and $23$ features on a blind test database composed of $25$ MRIs (i.e. not used in training phase), in the algorithm cited in \cite{tangaro2013}, (see Tables \ref{sfs_features} and \ref{sbe_features} respectively).

By analyzing the two subsets of selected features, it resulted that $13$ of the $23$ extracted by the SBE method are also present in the sample of $36$ features obtained by the SFS technique. Most of them are Haralick and Statistical features, except for the positional and Haar-like features, confirming the most importance given by Haralick and Statistical types as well as a very low contribution of Haar-like type.

We remark that, by minimizing the presence of Haralick features, in particular the correlations, it allows to improve the processing time as well as a better handling of the information content. In fact, among the three categories of features considered here, the Haralick type was the most \textit{time consuming} from the computational point of view.

The comparison of our FS methods with the widely used PCA demonstrates the very low performance of the PCA technique (as shown in Figure \ref{box_1}). This result is in agreement with the well-known downside of the method in presence of a very high non-linearity of the feature correlations. It is also an indirect confirmation about the intrinsic difficulty to separate the \textit{hippocampus} vs \textit{not-hippocampus} classes from MRI images.

We conclude that the SFS and SBE techniques are two promising methods allowing to reduce the input space size, with a very low loss of information and permitting classification performances comparable or even better than the case with a larger amount of features.

In fact, in terms of feature space dimension comparison, Morra \cite{morra2008} performs a voxel-based segmentation using about $18,000$ features with the weighted voting method AdaBoost \cite{freund1997} tested on a different image data set. In addition, FreeSurfer \cite{freesurfer2012}, which is a not voxel-based method considered a standard benchmark for MRI segmentation experiments, reaches a Dice value of $0.76 \pm 0.05$.

In this work, we observed that the selected features from both SFS and SBE methods are related to the high frequency component of the image. So this result would suggest which kind of features are best suitable for high frequency classification problems such as edge recognition. In fact, these correlation features, being based on intensity differences, are able
to capture local information based on discontinuity rather than similarity.

Besides, this result is a further suggestion for a future investigation: to put in practice a preprocessing procedure to enhance the contours of the structures contained in the image and to assess the usefulness of these procedures in the diagnosis support systems.

\section*{Acknowledgments}
The authors would like to thank the anonymous referee for extremely valuable comments and suggestions.




\bibliographystyle{splncs}
\bibliographystyle{elsarticle-num}

\end{document}